\documentclass[letter]{aa}
\usepackage{graphicx}
\usepackage[authoryear]{natbib}
\usepackage{amsmath}
\usepackage{amssymb}
\usepackage{txfonts}
\usepackage{threeparttable}
\usepackage[final]{microtype}
\begin{document}
   \title{Outburst of GX\,304$-$1 monitored with INTEGRAL: Positive correlation
   between the cyclotron line energy and flux}

   \author{D.~Klochkov\inst{1}
          \and
          V.~Doroshenko\inst{1}
          \and
          A.~Santangelo\inst{1}
          \and
          R.~Staubert\inst{1}
          \and
          C.~Ferrigno\inst{2}
          \and
          P.~Kretschmar\inst{3}
          \and
          I.~Caballero\inst{4}
          \and
          J.~Wilms\inst{5}
          \and
          I.~Kreykenbohm\inst{5}
          \and
          K.~Pottschmidt\inst{6,9}
          \and
          R.\,E.~Rothschild\inst{7}
          \and
          C.\,A.~Wilson-Hodge\inst{8}
          \and
          G.\,P\"uhlhofer\inst{1}
          }

   \institute{Institut f\"ur Astronomie und Astrophysik, Universit\"at
     T\"ubingen (IAAT), Sand 1, 72076 T\"ubingen, Germany
     \and
     ISDC Data Center for Astrophysics of the University of Geneva
     chemin d'\'Ecogia, 16 1290 Versoix, Switzerland
     \and
     European Space Astronomy Centre (ESA/ESAC), Science Operations
     Department, Villanueva de la Can\~ada (Madrid), Spain
     \and
     CEA Saclay, DSM/IRFU/SAp - UMR AIM (7158) CNRS/CEA/Universit\'e P. Diderot, Orme des Merisiers, B\^at. 709, 91191 Gif-sur-Yvette, France
     \and
     Dr. Karl Remeis Sternwarte \& Erlangen Centre for Astroparticle Physics, Sternwartstr. 7, 96049 Bamberg, Germany
     \and
     Center for Space Science and Technology, University of Maryland Baltimore County, 1000 Hilltop Circle, Baltimore, MD 21250, USA 
     \and
     Center for Astrophysics and Space Sciences, University of California, San Diego, 9500 Gilman Dr., La Jolla, CA 92093-0424, USA
     \and
     NASA Marshall Space Flight Center, Huntsville, AL 35812, USA
     \and
     CRESST \& NASA Goddard Space Flight Center, Astrophysics Science Division, Code 661, Greenbelt, MD 20771, USA
   }
   
   \date{Received ***, 2012; accepted ***, 2012}

  \abstract
   {X-ray spectra of many accreting pulsars exhibit significant 
     variations as a function of flux and thus of mass accretion rate. 
     In some of these pulsars,
     the centroid energy of the cyclotron line(s), which characterizes
     the magnetic field strength at the site of the X-ray emission, has 
     been found to vary systematically with flux. }
   {GX\,304$-$1 is a recently established cyclotron line source with a
     line energy around 50 keV. Since 2009, the pulsar shows
     regular outbursts with the peak flux exceeding one Crab. We
     analyze the \textit{INTEGRAL} observations of the source during its 
     outburst in January--February 2012.}
   {The observations covered almost the entire outburst, allowing us to measure
     the source's broad-band X-ray spectrum at different flux levels.
     We report on the variations in the spectral parameters with luminosity
     and focus on the variations in the cyclotron line.}
   {The centroid energy of the line is found to be positively correlated
     with the luminosity. We interpret this result as a manifestation
     of the local sub-Eddington (sub-critical) accretion regime operating in
     the source.}
   {}

   \titlerunning{Cyclotron line in GX\,304$-$1 correlates positively with flux}
   \maketitle
%
\section{Introduction}

In accreting binary pulsars, matter from the normal stellar companion 
is transferred to a highly magnetized ($B\gtrsim$10$^{12}$\,G) neutron
star. In the vicinity of the accretor, the gas flow is 
disrupted by the neutron star's magnetic field and channeled towards
the magnetic poles, where most of the X-rays originate.
The physics and the structure of the X-ray emitting region(s) above the 
neutron star surface 
are still highly debated
\citep[see, e.g.,][]{Becker:Wolff:07,Farinelli:etal:12}.
Since the matter hitting the accretor's surface is highly ionized, the 
magnetic field strength is a crucial parameter determining 
the physical processes inside the emitting region and the 
formation of the observed X-ray spectrum.
A direct way to assess the $B$-field strength at the site of X-ray emission
is the measurement of the cyclotron resonant scattering
features (CRSF or \emph{cyclotron lines}) in the X-ray spectrum of a 
pulsar. These features appear as absorption lines, caused by the 
resonant scattering
of photons off the electrons in Landau levels
\citep[see, e.g.,][]{Truemper:etal:78,Isenberg:etal:98}.
The energy of the fundamental line and the spacing between the
harmonics are directly proportional to the field strength.

In some accreting pulsars, the energy of the cyclotron line has been found to 
vary with luminosity, apparently due to a displacement of the line 
formation region. Such variations of the line energy
have been reported for V\,0332+53 \citep[e.g.,][]{Tsygankov:etal:10},
4U\,0115+63 \citep[e.g.,][]{Tsygankov:etal:10}, \hbox{Her\,X-1}
\citep{Staubert:etal:07, Klochkov:etal:11}, and 
A\,0535+26 \citep{Klochkov:etal:11}. The luminosity-dependence of
the cyclotron feature has strong implications for the physics of 
the X-ray emitting region as discussed, e.g., by \citet{Staubert:etal:07},
\citet{Klochkov:etal:11}, and \citet{Becker:Wolff:07}.

GX\,304$-$1 is a recently established cyclotron line source 
\citep{Yamamoto:etal:11}. It was discovered in a balloon
experiment in 1967 and subsequently identified as an X-ray pulsar
with a period of $\sim$272\,s \citep{McClintock:etal:77}. The
system contains a Be-type optical companion \citep{Mason:etal:78} and is
located at a distance of $\sim$2.4\,kpc \citep{Parkes:etal:80}. 
Since 1980, GX\,304$-$1 has remained in a quiescent state,
showing no outbursts. Starting from 2008, when the source was
detected with \textsl{INTEGRAL} \citep{Manousakis:etal:08}, GX\,304$-$1
``resumed'' its activity exhibiting outbursts with a period of
$\sim$132.5\,d. 

The energy of the cyclotron line in GX\,304$-$1 was measured
with \emph{Suzaku} and \textsl{RXTE} to be around 52\,keV 
by \citet{Yamamoto:etal:11}.
These authors demonstrate that the data taken at different flux levels show an 
indication of a positive correlation between the line energy and 
the X-ray flux, 
although at a low significance level. This made GX\,304$-$1 a good target 
for a luminosity-dependent study of the cyclotron feature.
In this work, we present the analysis of \textsl{INTEGRAL} data 
(see next section for the mission description)
taken during
the January 2012 outburst of the source. The data reveal a positive 
correlation between the cyclotron line energy and the source flux,
as well as the variation in other spectral parameters during the outburst.
We discuss our finding in the context of a model assuming that
different accretion regimes can operate 
in a particular pulsar depending on its X-ray luminosity.

\section{Observations and data reduction}

At the beginning of January 2012, GX\,304$-$1 entered an outburst
as reported by \citet{Yamamoto:etal:12}.
The outburst was monitored
by the \textit{International Gamma-Ray Astrophysics Laboratory} -
\textsl{INTEGRAL} \citep{Winkler:etal:03}, starting at MJD $\sim$55943.5,
when the source flux in the $\sim$20--80\,keV range was $\sim$250\,mCrab,
through the maximum of the outburst, when the source flux exceeded
one Crab, to MJD $\sim$55965.5, when the flux dropped to $\sim$100 mCrab.
\textsl{INTEGRAL} performed one observation 
every satellite orbit (about three days),
with a typical exposure of a few tens of kiloseconds each.
In total, eight observations were performed. 
The \textsl{INTEGRAL} scientific payload contains three X-ray
instruments: (i) the imager IBIS sensitive from $\sim$20\,keV to a
few MeV 
\citep{Ubertini:etal:03,Lebrun:etal:03}; 
(ii) the spectrometer SPI sensitive in roughly the same energy range as IBIS
\citep{Vedrenne:etal:03}; and (iii) the X-ray monitor \hbox{JEM-X} 
operating between $\sim$3 and $\sim$35\,keV \citep{Lund:etal:03}. 
Table\,\ref{tab:obs} summarizes the
\textsl{INTEGRAL} observations of GX\,304$-$1.
The increased solar activity during the observations
led to the reduction in the exposure time and availability of
the instruments as can be seen from the exposure columns of 
Table\,\ref{tab:obs}.

\begin{table}
  \centering
  \caption{\textsl{INTEGRAL} observations of the GX\,304$-$1 outburst in 
    January--February 2012.}
  \label{tab:obs}
  \vspace*{1mm}
  \begin{tabular}{l l l l l l}
    \hline\hline
    Rev.   &  Obs. ID   & Mid. MJD    & \multicolumn{3}{c}{Exposure [ksec]}  \\
           &            &             & JEM-X       &  IBIS          & SPI    \\
    \hline
    1131   & 09400230006 & 55944.0     & 64.6        &    42.7       & 68.6   \\
    1132   & 09400230007 & 55947.0     & 42.4        &    31.9       & 36.6   \\
    1133   & 09400230008 & 55950.0     & --          &    --         & 10.7   \\
    1134   & 09400230009 & 55952.8     & 7.3         &    25.4       & 37.8   \\
    1135   & 09400230010 & 55955.7     & --          &    6.7        & 25.1   \\
    1136   & 09400230011 & 55958.7     & 36.9        &    28.1       & 32.9   \\
    1137   & 09400230012 & 55962.0     & 78.1        &    59.7       & 78.4   \\
    1138   & 09400230013 & 55965.0     & 60.7        &    45.2       & 52.3   \\
    \hline
  \end{tabular}
\end{table}

\begin{figure}
\centering
\resizebox{0.95\hsize}{!}{\includegraphics{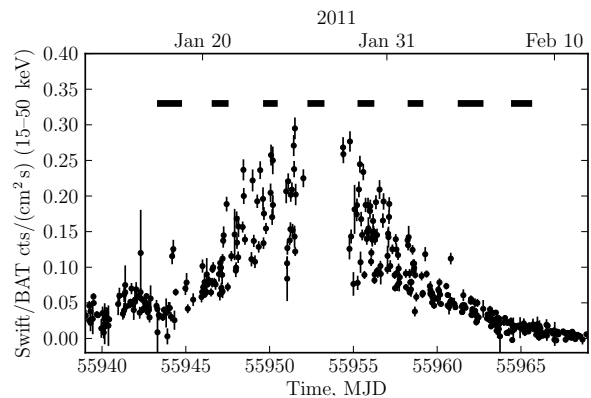}}
\caption{\textsl{INTEGRAL} observations of GX\,304$-$1 (horizontal bars)
superposed on the \textsl{Swift}/BAT light curve of the source during 
its outburst in January--February 2012 ($\sim$0.22 units of the vertical axis
corresponds to one Crab). 
}
\label{fig:lcobs}
\end{figure}

The \textsl{INTEGRAL} observations are indicated in Fig\,\ref{fig:lcobs},
which shows the \textsl{Swift}/BAT light curve of 
GX\,304$-$1\footnote{We used the \textsl{Swift}/BAT
transient monitor results provided by the \textsl{Swift}/BAT team}. 
The \textsl{INTEGRAL} monitoring has an excellent coverage of the 
entire outburst, providing a rare opportunity to follow the 
evolution of the outburst 
from the early rising phase to the late decay phase.

For our analysis, we used data from the ISGRI detector 
of IBIS, which is sensitive in the
20--300\,keV energy range, JEM-X, and SPI.
Standard data processing was performed with
version 9 of the Offline Science Analysis (OSA) software.
We performed an
additional gain correction of the ISGRI energy scale based
on the background spectral lines of Tungsten (the modified response
files based on the nearest Crab observations were used).

\section{Spectral analysis}

Figure \ref{fig:lcobs} shows that the X-ray flux of GX\,304$-$1
during the observed part of the outburst changed by an order of
magnitude. This allowed a detailed study of the 
luminosity dependence of the source's broad-band X-ray spectrum.
X-ray pulsations with a period of $\sim$274.9\,s were detected
in all \textsl{INTEGRAL} observations. This value is roughly consistent
with the known pulse period. A detailed timing analysis
is not  part of the present work and will be presented elsewhere. 

\begin{figure}
\centering
\resizebox{0.90\hsize}{!}{\includegraphics{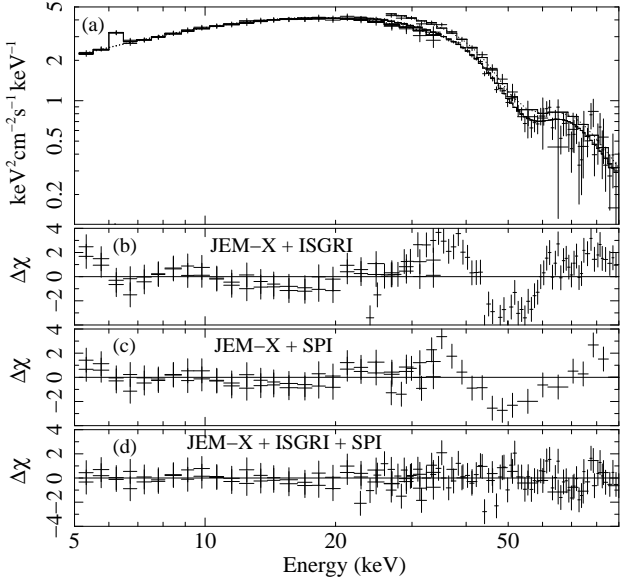}}
\caption{Example of the  \textsl{INTEGRAL} spectrum of GX\,304$-$1 
(revolution 1132) with a simultaneous fit of data from JEM-X, ISGRI, 
and SPI with a power-law/cutoff model including a cyclotron absorption line
(see text) (a), the residuals for a fit with the model without the cyclotron 
line for JEM-X + ISGRI (b), and JEM-X + SPI (c), 
and with the model where the line is included for data from all instruments (d).
}
\label{fig:spe}
\end{figure}

In all observations, the X-ray continuum could be closely
modeled by a standard power-law/cutoff function 
(flux $\propto E^{-\Gamma}\exp{[E/E_{\rm fold}]}$, where $E$ is the 
photon energy, $\Gamma$ is the photon index, and $E_{\rm fold}$ is
the exponential roll-off parameter)
modified by photo-electric absorption at low energies. 
In addition, the spectra showed a cyclotron resonant
scattering feature around $\sim$50\,keV in absorption, which was
modeled with a multiplicative absorption line with a Gaussian
optical depth profile
$I(E)=
I_{\rm cont}(E)\cdot e^{-G(E)}$, where $G(E) = -\tau_{\rm cyc}/(\sqrt{2\pi}\sigma_{\rm cyc})
\cdot\exp[-0.5(E-E_{\rm cyc})^2/\sigma_{\rm cyc}^2]$,
$I_{\rm cont}(E)$ is the continuum function,
$E_{\rm cyc}$, $\sigma_{\rm cyc}$, and $\tau_{\rm cyc}$ are 
the centroid energy, width, and optical depth of the line,
respectively. The line is clearly detected in
ISGRI and SPI data separately, as shown by the residual
plots in Fig.\,\ref{fig:spe}. 
The inclusion of the absorption line in the model leads
to an improvement in the reduced $\chi^2$ from 3.10 for 145 d.o.f. 
(with the corresponding null-hypothesis probability
of only $\sim$10$^{-32}$)
to 0.75 for 142 d.o.f. 
(null-hypothesis probability $>$0.9).
The energy of the line
is consistent with that reported by \citet{Yamamoto:etal:11}
based on the \textsl{RXTE} and \emph{Suzaku} observations. 
We also included an additive Gaussian component to model 
the Fe $K_\alpha$ fluorescence emission line around 6.4\,keV. 
The inclusion of the line reduces the
residuals around 6\,keV. The corresponding improvement of the 
reduced $\chi^2$ is, however, marginal: from, 0.80 (143 d.o.f.) to 
0.75 (142 d.o.f.). The presence of the Fe line is, therefore, questionable.

Both the continuum and cyclotron line parameters vary systematically
during the outburst. Here, we focus on the evolution of the 
cyclotron line energy $E_{\rm cyc}$, to
establish a possible correlation of $E_{\rm cyc}$ with flux,
similar to that found by \citet{Yamamoto:etal:11}.
To characterize
this variability, we used only observations where data from 
all three  \textsl{INTEGRAL} X-ray 
instruments were available. This filtering allowed
us to obtain a homogeneous set of data and, hence minimize possible
systematic effects. Following this selection, data from the orbits 1133 
and 1135 were excluded. 
The JEM-X data collected during the orbit 1134 
(peak of the outburst) have been
flagged as ``bad'' by the instrument team owing to the impossibility
to provide a precise energy calibration. In our analysis, however, 
we use the JEM-X data to restrict the low-energy part of 
the broad-band X-ray continuum of the source, for which a precise energy
scale is not very critical.
Therefore, after consulting the JEM-X team, we re-introduced the JEM-X
data from the revolution 1134 in our analysis. We checked, however,
that inclusion of the JEM-X data of rev. 1134 to the 
corresponding spectral fit does not significantly influence
the measured energy of the cyclotron line (the focus of this work),
but only affects its uncertainty.
Table\,\ref{tab:par} summarizes the best-fit
spectral parameters achieved in each observations. 

\begin{table*}
\caption{Best-fit spectral parameters with the corresponding 
1$\sigma$-uncertainties for the \textit{INTEGRAL} observations used in this work.
The luminosity is provided assuming a distance of 2.4\,kpc and an isotropic
emission diagram.}
\label{tab:par}
\centering
\renewcommand{\arraystretch}{1.3}\begin{tabular}{l | c c c c c c c}
\hline\hline
Revolution & 1131& 1132& 1134& 1136& 1137& 1138\\
\hline
$\Gamma$ & $1.10_{-0.03}^{+0.03}$ & $0.83_{-0.03}^{+0.03}$ & $0.93_{-0.09}^{+0.09}$ & $1.10_{-0.08}^{+0.08}$ & $1.23_{-0.08}^{+0.08}$ & $1.56_{-0.05}^{+0.13}$\\
$E_{\rm fold}$, keV & $19.1_{-0.6}^{+0.7}$ & $15.5_{-0.4}^{+0.4}$ & $14.5_{-0.6}^{+0.6}$ & $16.9_{-0.7}^{+0.8}$ & $19.5_{-1.0}^{+1.1}$ & $28.6_{-2.2}^{+3.8}$\\
$E_{\rm cyc}$, keV & $52.98_{-0.75}^{+0.80}$ & $54.39_{-0.73}^{+0.75}$ & $55.28_{-0.81}^{+0.87}$ & $51.44_{-0.52}^{+0.54}$ & $51.70_{-0.63}^{+0.66}$ & $48.37_{-1.12}^{+1.19}$\\
$\sigma_{\rm cyc}$, keV & $6.45_{-0.58}^{+0.62}$ & $7.56_{-0.57}^{+0.59}$ & $8.45_{-0.75}^{+0.83}$ & $5.60_{-0.46}^{+0.48}$ & $5.44_{-0.55}^{+0.56}$ & $4.80_{-0.89}^{+0.91}$\\
$\tau_{\rm cyc}$ & $9.4_{-1.2}^{+1.3}$ & $12.1_{-1.3}^{+1.4}$ & $13.3_{-2.0}^{+2.4}$ & $8.8_{-0.9}^{+0.9}$ & $8.9_{-1.1}^{+1.2}$ & $5.7_{-1.3}^{+1.3}$ \\
Flux/$10^{-8}$\,erg\,s$^{-1}$\,cm$^{-2}$ & $0.375_{-0.002}^{+0.002}$ & $0.811_{-0.004}^{+0.004}$ & $1.630_{-0.011}^{+0.011}$ & $0.729_{-0.005}^{+0.005}$ & $0.340_{-0.003}^{+0.002}$ & $0.151_{-0.002}^{+0.002}$\\
$L_{\rm X}$/10$^{37}$\,erg~s$^{-1}$ & 0.26 & 0.56 & 1.12 & 0.50 & 0.23 & 0.10 \\
$\chi^2_{\rm red}$/d.o.f. & 1.1/164 & 0.7/164 & 1.3/164 & 1.1/164 & 0.9/164 & 1.0/164\\
\hline
\end{tabular}
\end{table*}

\begin{figure}
\centering
\resizebox{0.92\hsize}{!}{\includegraphics{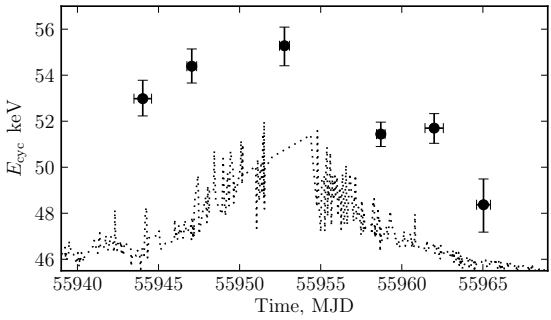}}
\caption{Evolution of the cyclotron line centroid energy $E_{\rm cyc}$
throughout the outburst of GX\,304$-$1 as measured with \textsl{INTEGRAL}. 
The vertical error bars indicate 1$\sigma$-uncertainties. 
The horizontal error bars indicate the time intervals of the observations. 
The dotted line represents the re-scaled Swift/BAT light curve. 
}
\label{fig:ecyclc}
\end{figure}

\begin{figure}
\centering
\resizebox{0.80\hsize}{!}{\includegraphics{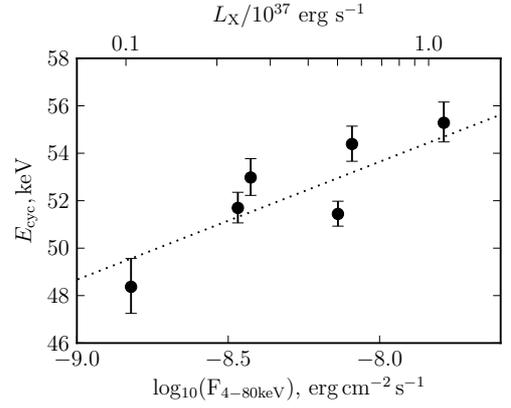}}
\caption{Cyclotron line centroid energy $E_{\rm cyc}$ as a function
of the logarithm of flux in the 4--80\,keV range.
The error bars indicate 1$\sigma$-uncertainties
(the flux uncertainties are smaller than the symbol size).
The top X-axis represents the corresponding 
isotropic source luminosity
assuming a distance of 2.4\,kpc.
The dotted line shows a linear fit to the $E_{\rm cyc}$--log$_{10}$(Flux)
dependence.
}
\label{fig:ecyc}
\end{figure}

A clear systematic variation in the line energy over the outburst is evident in 
Fig.\,\ref{fig:ecyclc}, which shows that $E_{\rm cyc}$ generally follows
the X-ray flux.  
To assess the interdependence of the two parameters, we
plotted $E_{\rm cyc}$ as a function of the X-ray flux in the 4--80\,keV
range measured with \textsl{INTEGRAL} in the respective observations
(Fig.\,\ref{fig:ecyc}). The plot 
shows a positive correlation between the two values.
A linear fit to the dependence of $E_{\rm cyc}$ on
the logarithm of flux (dotted line in the plot)
reveals a slope of 4.97$\pm$1.12 
keV/log$_{10}$(erg\,cm$^{-2}$\,s$^{-1}$).
The standard linear correlation analysis of the 
$E_{\rm cyc}$--log$_{10}$(Flux) 
dependence yields 
a Pearson's correlation coefficient of 0.88 with
a probability of obtaining the correlation by chance 
of $\sim$0.01 (one-sided).

\begin{figure}
\centering
\resizebox{0.90\hsize}{!}{\includegraphics{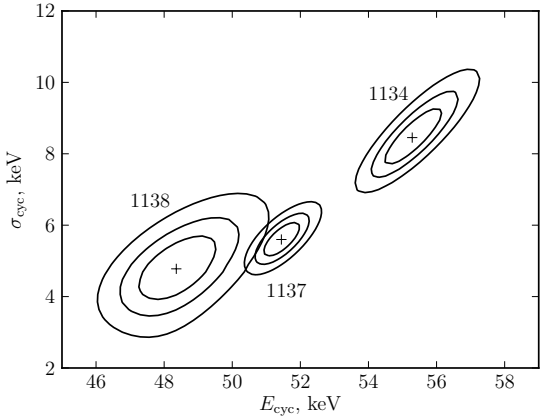}}
\caption{$\chi^2$-contour plots of the parameter pair 
$E_{\rm cyc}$/$\sigma_{\rm cyc}$ for a few selected 
observations. The contours correspond to $\chi^2_{\rm min}$+1.0
(the projections of this contour to the parameter axes correspond
to the 68\%-uncertainty for one parameter of interest), 
$\chi^2_{\rm min}$+2.3 (68\%-uncertainty for two parameters of
interest), and  $\chi^2_{\rm min}$+4.61 (90\%-uncertainty for 
two parameters of interest). The respective orbit numbers
are indicated.
}
\label{fig:con}
\end{figure}

We verified whether our spectral fits contain artificial (model-driven)
dependences of the cyclotron line energy $E_{\rm cyc}$ 
and other model parameters using $\chi^2$-contour plots. No significant 
model-driven dependences between $E_{\rm cyc}$ and any of the 
continuum parameters were found. The line energy was, however, found to be 
somewhat coupled
to the line width $\sigma_{\rm cyc}$ and its central optical depth 
$\tau_{\rm cyc}$. Nevertheless, the $\chi^2$-minima and the confidence 
intervals could be clearly identified and are separated for the 
different observations as 
shown in Fig.\,\ref{fig:con}. The plot shows the contours
for the $E_{\rm cyc}$/$\sigma_{\rm cyc}$ pair. The  
$E_{\rm cyc}$/$\tau_{\rm cyc}$ contours look similar.

To check whether the $E_{\rm cyc}$/flux correlation is related to
instrumental effects, we performed spectral fits using  
only the JEM-X and SPI data (excluding ISGRI). To verify whether
the correlation depends on the choice of the spectral model, we
fit the data using a Lorentzian line profile instead of a Gaussian
one. We also tried two alternative continuum functions: XSPEC 
\texttt{powerlaw$\times$highecut} and \texttt{compTT} models.
In the former model, the \texttt{highecut} component controls
the exponential roll-off. In addition to $E_{\rm fold}$, this 
component has an additional
parameter -- the cutoff energy $E_{\rm cutoff}$, above which the spectrum
is affected by the roll-off. In our fits, $E_{\rm cutoff}$ stays between a few
and $\sim$10\,keV.
In all cases, the positive $E_{\rm cyc}$/flux 
correlation was reproduced.
We conclude, therefore, that the reported correlation 
arises from the source's behavior and reflects real physics.

\section{Discussion and conclusions}

The positive correlation between the cyclotron line 
centroid energy and the flux  found with \textsl{INTEGRAL} confirms the 
claim of this correlation by \citet{Yamamoto:etal:11}.
The revealed dependence 
establishes GX\,304$-$1 as the third member of a
slowly emerging 
class of accreting pulsars showing a positive $E_{\rm cyc}$/flux 
correlation with the other members being Her\,X-1 \citep{Staubert:etal:07} 
and possibly A\,0535+26 \citep{Klochkov:etal:11}, for which the positive
correlation has so far been only established in the pulse-to-pulse analysis.
The opposite (negative) correlation between the line energy and the flux was 
found 
in V\,0332+53 and 4U\,0115+63 
\citep[e.g.,][]{Tsygankov:etal:07, Tsygankov:etal:10}.
According to discussions in \citet{Staubert:etal:07},
\citet{Klochkov:etal:11}, and
\citet{Becker:etal:12}, these two types of dependences reflect two different
regimes of accretion. A particular regime is realized in a source depending
on whether its X-ray luminosity $L$ is above or below a critical 
luminosity $L_\mathrm{c}$, which corresponds to the local Eddington luminosity at
the X-ray emitting structure(s) on/above the neutron star surface.
In accreting pulsars radiating above $L_\mathrm{c}$  
(``super-critical'' sources), 
infalling matter is decelerated in a radiative shock, 
whose height is believed to increase with $L$, i.e., drift
towards an area with a 
lower $B$-field strength. The opposite behavior probably 
occurs in sources radiating below or close to $L_\mathrm{c}$ 
(``sub-critical'' sources), 
where infalling matter is stopped by the Coulomb drag and
collective plasma effects rather than in a radiative shock.
As discussed in \citet{Staubert:etal:07} and 
\citet{Becker:etal:12}, the height of the emitting region 
decreases with increasing luminosity owing to a corresponding
increase in ram pressure of the infalling material, 
leading to a positive $E_{\rm cyc}$/flux correlation.
The critical luminosity $L_\mathrm{c}$ depends on the parameters
of the accreting neutron star, but is generally
around a few times $\sim$10$^{37}$\,erg\,s$^{-1}$ 
\citep{Basko:Sunyaev:76,Staubert:etal:07,Becker:etal:12}. 
Assuming a distance of 2.4\,kpc \citep{Parkes:etal:80}, the X-ray luminosity of 
GX\,304$-$1 in the 4--80\,keV range during the reported \textsl{INTEGRAL}
observations varies between $\sim$1.1$\times 10^{36}$\,erg\,s$^{-1}$ and 
$\sim$1.13$\times 10^{37}$\,erg\,s$^{-1}$. Thus, according to the 
described picture, the source should belong to the class of 
``sub-critical'' sources, for which a positive $E_{\rm cyc}$/flux correlation
is expected. The reported observations are, therefore,
in agreement with the idea of two accretion regimes and increases the
yet very small sample of accreting pulsars with established
$E_{\rm cyc}$/flux correlations.

\begin{acknowledgements}
The work was supported by the Carl-Zeiss-Stiftung.
JW and IK were partially supported by BMWi under DLR grant 50 OR 1007.
This research is based on observations with \textsl{INTEGRAL}, an ESA
project with instruments and science data centre funded by ESA member
states.
We thank the \textsl{INTEGRAL} team for the prompt
scheduling of the TOO observations and support
with the data reduction and calibration.
We thank ISSI (Bern, Switzerland) for its hospitality during the
collaboration meetings of our team.
\end{acknowledgements}

\bibliographystyle{aa}
\bibliography{ref}

\begin{thebibliography}{23}
\expandafter\ifx\csname natexlab\endcsname\relax\def\natexlab#1{#1}\fi

\bibitem[{{Basko} \& {Sunyaev}(1976)}]{Basko:Sunyaev:76}
{Basko}, M.~M. \& {Sunyaev}, R.~A. 1976, \mnras, 175, 395

\bibitem[{Becker {et~al.}(2012)Becker, {Klochkov}, Schoenherr, Nishimura,
  {Ferrigno}, Caballero, Kretschmar, T., Wilms, \& {Staubert}}]{Becker:etal:12}
Becker, P., {Klochkov}, D., Schoenherr, G., {et~al.} 2012, \aap, submitted

\bibitem[{{Becker} \& {Wolff}(2007)}]{Becker:Wolff:07}
{Becker}, P.~A. \& {Wolff}, M.~T. 2007, \apj, 654, 435

\bibitem[{{Farinelli} {et~al.}(2012){Farinelli}, {Ceccobello}, {Romano}, \&
  {Titarchuk}}]{Farinelli:etal:12}
{Farinelli}, R., {Ceccobello}, C., {Romano}, P., \& {Titarchuk}, L. 2012, \aap,
  538, A67

\bibitem[{{Isenberg} {et~al.}(1998){Isenberg}, {Lamb}, \&
  {Wang}}]{Isenberg:etal:98}
{Isenberg}, M., {Lamb}, D.~Q., \& {Wang}, J.~C.~L. 1998, \apj, 493, 154

\bibitem[{{Klochkov} {et~al.}(2011){Klochkov}, {Staubert}, {Santangelo},
  {Rothschild}, \& {Ferrigno}}]{Klochkov:etal:11}
{Klochkov}, D., {Staubert}, R., {Santangelo}, A., {Rothschild}, R.~E., \&
  {Ferrigno}, C. 2011, \aap, 532, A126

\bibitem[{{Lebrun} {et~al.}(2003){Lebrun}, {Leray}, {Lavocat}, {Cr{\'e}tolle},
  {Arqu{\`e}s}, {Blondel}, {Bonnin}, {Bou{\`e}re}, {Cara}, {Chaleil}, {Daly},
  {Desages}, {Dzitko}, {Horeau}, {Laurent}, {Limousin}, {Mathy}, {Mauguen},
  {Meignier}, {Molini{\'e}}, {Poindron}, {Rouger}, {Sauvageon}, \&
  {Tourrette}}]{Lebrun:etal:03}
{Lebrun}, F., {Leray}, J.~P., {Lavocat}, P., {et~al.} 2003, \aap, 411, L141

\bibitem[{{Lund} {et~al.}(2003){Lund}, {Budtz-J{\o}rgensen}, {Westergaard},
  {Brandt}, {Rasmussen}, {Hornstrup}, {Oxborrow}, {Chenevez}, {Jensen},
  {Laursen}, {Andersen}, {Mogensen}, {Rasmussen}, {Om{\o}}, {Pedersen},
  {Polny}, {Andersson}, {Andersson}, {K{\"a}m{\"a}r{\"a}inen}, {Vilhu},
  {Huovelin}, {Maisala}, {Morawski}, {Juchnikowski}, {Costa}, {Feroci},
  {Rubini}, {Rapisarda}, {Morelli}, {Carassiti}, {Frontera}, {Pelliciari},
  {Loffredo}, {Mart{\'{\i}}nez N{\'u}{\~n}ez}, {Velasco}, {Larsson},
  {Svensson}, {Zdziarski}, {Castro-Tirado}, {Attina}, {Goria}, {Giulianelli},
  {Cordero}, {Rezazad}, {Schmidt}, {Carli}, {Gomez}, {Jensen}, {Sarri},
  {Tiemon}, {Orr}, {Much}, {Kretschmar}, \& {Schnopper}}]{Lund:etal:03}
{Lund}, N., {Budtz-J{\o}rgensen}, C., {Westergaard}, N.~J., {et~al.} 2003,
  \aap, 411, L231

\bibitem[{{Manousakis} {et~al.}(2008){Manousakis}, {Beckmann}, {Bianchin},
  {Brandt}, {Chenevez}, {Hermsen}, {von Kienlin}, {Krivonos}, {Mas-Hesse},
  {Parmar}, \& {Reglero}}]{Manousakis:etal:08}
{Manousakis}, A., {Beckmann}, V., {Bianchin}, V., {et~al.} 2008, ATEL 1613

\bibitem[{{Mason} {et~al.}(1978){Mason}, {Murdin}, {Parkes}, \&
  {Visvanathan}}]{Mason:etal:78}
{Mason}, K.~O., {Murdin}, P.~G., {Parkes}, G.~E., \& {Visvanathan}, N. 1978,
  \mnras, 184, 45p

\bibitem[{{McClintock} {et~al.}(1977){McClintock}, {Nugent}, {Li}, \&
  {Rappaport}}]{McClintock:etal:77}
{McClintock}, J.~E., {Nugent}, J.~J., {Li}, F.~K., \& {Rappaport}, S.~A. 1977,
  \apjl, 216, L15

\bibitem[{{Parkes} {et~al.}(1980){Parkes}, {Murdin}, \&
  {Mason}}]{Parkes:etal:80}
{Parkes}, G.~E., {Murdin}, P.~G., \& {Mason}, K.~O. 1980, \mnras, 190, 537

\bibitem[{{Priedhorsky} \& {Terrell}(1983)}]{Priedhorsky:Terrell:83}
{Priedhorsky}, W.~C. \& {Terrell}, J. 1983, \apj, 273, 709

\bibitem[{{Protassov} {et~al.}(2002){Protassov}, {van Dyk}, {Connors},
  {Kashyap}, \& {Siemiginowska}}]{Protassov_etal02}
{Protassov}, R., {van Dyk}, D.~A., {Connors}, A., {Kashyap}, V.~L., \&
  {Siemiginowska}, A. 2002, ApJ, 571, 545

\bibitem[{{Staubert} {et~al.}(2007){Staubert}, {Shakura}, {Postnov}, {Wilms},
  {Coburn}, {Rodina}, \& {Klochkov}}]{Staubert:etal:07}
{Staubert}, R., {Shakura}, N.~I., {Postnov}, K., {et~al.} 2007, \aap, 465, L25

\bibitem[{{Tr\"umper} {et~al.}(1978){Tr\"umper}, {Pietsch}, {Reppin}, {Voges},
  {Staubert}, \& {Kendziorra}}]{Truemper:etal:78}
{Tr\"umper}, J., {Pietsch}, W., {Reppin}, C., {et~al.} 1978, \apjl, 219, L105

\bibitem[{{Tsygankov} {et~al.}(2007){Tsygankov}, {Lutovinov}, {Churazov}, \&
  {Sunyaev}}]{Tsygankov:etal:07}
{Tsygankov}, S.~S., {Lutovinov}, A.~A., {Churazov}, E.~M., \& {Sunyaev}, R.~A.
  2007, Astronomy Letters, 33, 368

\bibitem[{{Tsygankov} {et~al.}(2010){Tsygankov}, {Lutovinov}, \&
  {Serber}}]{Tsygankov:etal:10}
{Tsygankov}, S.~S., {Lutovinov}, A.~A., \& {Serber}, A.~V. 2010, \mnras, 401,
  1628

\bibitem[{{Ubertini} {et~al.}(2003){Ubertini}, {Lebrun}, {Di Cocco}, {Bazzano},
  {Bird}, {Broenstad}, {Goldwurm}, {La Rosa}, {Labanti}, {Laurent}, {Mirabel},
  {Quadrini}, {Reglero}, {Sabau}, {Sacco}, {Staubert}, {Vigroux}, {Weisskopf},
  \& {Zdziarski}}]{Ubertini:etal:03}
{Ubertini}, P., {Lebrun}, F., {Di Cocco}, G., {et~al.} 2003, \aap, 411, L131

\bibitem[{{Vedrenne} {et~al.}(2003){Vedrenne}, {Roques}, {Sch{\"o}nfelder},
  {Mandrou}, {Lichti}, {von Kienlin}, {Cordier}, {Schanne}, {Kn{\"o}dlseder},
  {Skinner}, {Jean}, {Sanchez}, {Caraveo}, {Teegarden}, {von Ballmoos},
  {Bouchet}, {Paul}, {Matteson}, {Boggs}, {Wunderer}, {Leleux},
  {Weidenspointner}, {Durouchoux}, {Diehl}, {Strong}, {Cass{\'e}}, {Clair}, \&
  {Andr{\'e}}}]{Vedrenne:etal:03}
{Vedrenne}, G., {Roques}, J.-P., {Sch{\"o}nfelder}, V., {et~al.} 2003, \aap,
  411, L63

\bibitem[{{Winkler} {et~al.}(2003){Winkler}, {Courvoisier}, {Di Cocco},
  {Gehrels}, {Gim{\'e}nez}, {Grebenev}, {Hermsen}, {Mas-Hesse}, {Lebrun},
  {Lund}, {Palumbo}, {Roques}, {Schnopper}, {Sch{\"o}nfelder}, {Sunyaev},
  {Teegarden}, {Ubertini}, {Vedrenne}, \& {Dean}}]{Winkler:etal:03}
{Winkler}, C., {Courvoisier}, T.~J.-L., {Di Cocco}, G., {et~al.} 2003, \aap,
  411, L1

\bibitem[{{Yamamoto} {et~al.}(2011){Yamamoto}, {Sugizaki}, {Mihara},
  {Nakajima}, {Yamaoka}, {Matsuoka}, {Morii}, \&
  {Makishima}}]{Yamamoto:etal:11}
{Yamamoto}, T., {Sugizaki}, M., {Mihara}, T., {et~al.} 2011, \pasj, 63, 751

\bibitem[{{Yamamoto} {et~al.}(2012){Yamamoto}, Tomida, Mihara, {Sugizaki},
  Serino, Nakahira, \& {et al.}}]{Yamamoto:etal:12}
{Yamamoto}, T., Tomida, H., Mihara, T., {et~al.} 2012, ATEL 3856

\end{thebibliography}

\end{document}